\documentstyle[11pt]{article}
\def\ft#1#2{{\textstyle{{\scriptstyle #1} \over {\scriptstyle #2}}}}
\def\st#1{{\scriptstyle #1}}

\def\ww3{{$W_3$}}

\def\del{\partial}

\begin{document}
\topmargin 0pt
\oddsidemargin 5mm
\begin{titlepage}
\begin{flushright}
CTP TAMU-13/95\\
hep-th/9503158\\
\end{flushright}
\vspace{1.5truecm}
\begin{center}
{\bf {\Large BRST Operators for Higher-spin Algebras}}
\footnote{Supported in part by the
U.S. Department of Energy, under grant DE-FG05-91-ER40633}
\vspace{1.5truecm}

{\large H. L\"u,  C.N. Pope and K.W. Xu}
\vspace{1.1truecm}

{\small Center for Theoretical Physics, Texas A\&M University,
                College Station, TX 77843-4242}

\end{center}

\vspace{1.0truecm}

\begin{abstract}
\vspace{1.0truecm}

    In this paper, we construct non-critical BRST operators for matter and
Liouville systems whose currents generate two different $W$ algebras.  At
the classical level, we construct the BRST operators for $W^{\rm
M}_{2,s}\otimes W^{\rm L}_{2,s'}$.  The construction is possible for $s=s'$
or $s\ge s'+2$.   We also obtain the BRST operator for $W^{\rm
M}_{2,4}\otimes W^{\rm L}_4$ at the classical level.  We use free scalar
realisations for the matter currents in the above constructions.   At the
full quantum level, we obtain the BRST operators for $W^{\rm M}_{2,s}\otimes
W^{\rm L}_2$ with $s=4,5, 6$, where $W_2$ denotes the Virasoro algebra.  For
the first and last cases, the BRST operators are expressed in terms of
abstract matter and Liouville currents.  As a by-product, we obtain the
$W_{2,4}$ algebra at $c=-24$ and the $W_{2,6}$ algebra at $c=-2$ and
$-\ft{286}3$, at which values the algebras were previously believed not to
exist.

\end{abstract}
\end{titlepage}
\newpage
\pagestyle{plain}

     Owing to the non-linearity of $W$ algebras, BRST methods seem to
provide the only reliable way of quantising the associated $W$--string
theories.   Since the construction of the critical BRST operator for the
$W_3$ algebra by Thierry-Mieg \cite{tm}, a number of developments have taken
place.  Critical BRST operators for more complicated $W$ algebras, including
$W_{2,s}$ for $s\le 8$ \cite{w2s}, $W_4$ \cite{horn,berg,zhu} and $W\!B_2$
\cite{zhu}, have been constructed.   In a remarkable development, it was
shown in Ref.\ \cite{blnw} that a non-critical $W_3$ BRST operator can be
constructed, involving currents generating two commuting copies of the $W_3$
algebra. One copy can be viewed as the matter sector, and the other as the
Liouville sector, with the sum of the central charges of the two sectors
adding up to the critical value.  Subsequently the non-critical $W_{2,4}$
BRST operator was obtained \cite{zhao}.  It is surprising that such
constructions can be carried out for systems with non-linear $W$ symmetries,
since one cannot simply add the currents of the two copies and obtain a
realisation of the algebra. This contrasts with the situation for a system
with a linear symmetry, where one merely adds the currents of the matter and
Liouville sectors to obtain a new critical realisation of the algebra.  For
a system with non-linear symmetry, however, the critical and non-critical
BRST operators are intrinsically different.

    The philosophy behind the construction of the non-critical BRST
operators discussed above is that there are two commuting copies of the same
$W$ algebra.  In this paper, we shall generalise the construction to the
case where the currents of the matter sector and Liouville sector generate
two different $W$ algebras.  The ghost fields of the BRST operator arise
from the gauge fixing of the local symmetries generated by the matter
currents.   Thus the $W$ algebra in the matter sector defines the ghost
content of the BRST operator.  We shall consider non-critical BRST operators
for the $W^{\rm M}_{2,s}\otimes W^{\rm L}_{2,s'}$, $W^{\rm M}_{2,4}\otimes
W^{\rm L}_4$ and $W^{\rm M}_{2,s} \otimes W^{\rm L}_2$ systems, where $W_2$
stands for the Virasoro algebra.

    To begin, we review some previous results for critical $W_{2,s}$
BRST operators.  The $W_{2,s}$ algebra exists at the classical level for all
positive integer values of $s$.  It is generated by the energy-momentum
tensor $T$ and a primary spin--$s$ current $W$.  The classical OPE
$W(z)W(0)$ is given by
\begin{equation}
W(z)W(w)\sim {2T^{s-1} \over (z-w)^2} + {\del T^{s-1} \over z-w}
\ .\label{wwope}
\end{equation}
The classical BRST operator is given by \cite{lptwx}
\begin{equation}
Q=\oint\Big( c(T -s\beta\del\gamma -(s-1)\del\beta\, \gamma -b\del c)
+\gamma\, W -\del\gamma\, \gamma b\, T^{s-2}\Big)\ ,\label{w2sbrstcr1}
\end{equation}
where $(c, b)$ and $(\gamma, \beta)$ are the ghosts and anti-ghosts for the
currents $T$ and $W$ respectively.   Unlike the case of the $W_3$ algebra,
this BRST operator cannot, for general values of $s$, be extended to a full
quantum BRST operator for the abstract $W_{2,s}$ algebra.  This is because
for general value of $s$, the $W_{2,s}$ algebra exists at the quantum level
only for certain discrete values of central charge, which do not include the
value needed for criticality.   However, it was shown \cite{w2s,lptwx} that
if one realises the classical currents with free scalars, the resulting
classical BRST operator {\it is} quantisable.  In this realisation, all
scalars but one distinguished scalar $\phi$ enter the currents only via
their energy-momentum tensor $T_X$ \cite{lptwx}:
\begin{eqnarray}
T&=& -\ft12 (\del\phi)^2 + T_X\ ,\nonumber\\
W&=& \sum_{n=0}^{[s/2]} s^{-1} (-2)^{-s/2}\, 2^{n+1} {s\choose 2n}
(\del\phi)^{s-2n}\, T^n_X\ .\label{w2sclreal}
\end{eqnarray}
In terms of this realisation, the BRST operator can be transformed by a
local canonical field redefinition into the following graded form
\cite{lptwx}:
\begin{eqnarray}
Q_0&=&\oint c\Big(T - s\, \beta\del\gamma -(s-1)\del\beta\, \gamma -b\del c
\Big)
\ ,\nonumber\\
Q_1&=&\oint \gamma\Big( (\del\phi)^s + \ft12 s^2\, (\del\phi)^{s-2}\,
\beta\del\gamma \Big)\ .\label{w2sbrstcr2}
\end{eqnarray}
To quantise the classical $W_{2,s}$ string and obtain a critical quantum
BRST operator, one can add $\hbar$--dependent quantum corrections to the
classical BRST operator.  Explicit solutions have been found for all $s\le
8$ \cite{w2s,lptwx}, and general existence arguments for one class of
solutions for all values of $s$ were given in Refs.\ \cite{west,w2s}.

    Now we turn to the construction of the non-critical BRST operators for
$W^{\rm M}_{2,s}\otimes W^{\rm L}_{2,s'}$.  Again we use the free-scalar
realisation (\ref{w2sclreal}) for the $W^{\rm M}_{2,s}$ system, so that the
corresponding BRST operator can be written in a graded form, generalising
Eqn.\ (\ref{w2sbrstcr2}).  By demanding nilpotence at the classical level,
we find that the BRST operator is given by $Q=Q_0 + Q_1$ with
\begin{eqnarray}
Q_0&=&\oint c\Big(T_{\phi} + T_X + T_{\rm L}- s\, \beta\del\gamma -
(s-1)\del\beta\, \gamma -b\del c\Big)
\ ,\nonumber\\
Q_1&=&\oint \gamma\Big( (\del\phi)^s +\sum_{n=1}^{[s'/2]} g_n(s')
(\del\phi)^{s-2n}\, T_{\rm L}^n + \sum_{n=1}^{[s'/2 +1]} h_n(s,s')
(\del\phi)^{s-2n}\, T_{\rm L}^{n-1}\,  \beta\del\gamma\nonumber\\
&& + 2^{-s'/2}\, s'\,(\del\phi)^{s-s'}\, W_{\rm L} +
2^{-1-s'/2}\, s'(s-s')^2\, (\del\phi)^{s-s'-2}\, W_{\rm L}\, \beta\del
\gamma\Big)\ ,\label{w2s2sp}
\end{eqnarray}
where the coefficients $g_n(s')$ and $h_n(s,s')$ are given by
\begin{eqnarray}
g_n(s')&=& {(-1)^n s'\, (s'-n-1)!\over 2^n\, n!\, (s'-2n)!}\ ,\nonumber\\
h_n(s,s')&=& {(-1)^{n+1} s'\, (s'-n-1)! \over
2^n\, (n-1)!\, (s'-2n+2)!}\Big ( s(s-4n+4)(s'-n) + s'(s'-2)(n-1)\Big)\ .
\label{coefghs}
\end{eqnarray}
The Liouville currents $T_{\rm L}$ and $W_{\rm L}$ satisfy the classical
$W_{2,s'}$ algebra, given by Eqn.\ (\ref{wwope}) with $s$ replaced by $s'$.
It follows from the last term in the $Q_1$ operator (\ref{w2s2sp}) that
either $s\ge s' + 2$ or $s=s'$.  The latter case corresponds to the
previously-discussed non-critical $W_{2,s}$ BRST operators.  The former case
corresponds to the new possibilities that we are considering in this paper.
When $s'=2$, the $W_{2,s'}$ algebra becomes a linear algebra, and the BRST
operator (\ref{coefghs}) is slightly modified:  The $W$ current in the
Liouville sector enters the $Q_1$ operator in the combination $W_{\rm L} -
T_{\rm L}$, which generates the Virasoro algebra.  Thus for the case
$s'=2$, the BRST operator can be simplified to the one for $W^{\rm
M}_{2,s}\otimes W^{\rm L}_2$, where $W_2$ denotes the Virasoro algebra.  We
find that it is given by
\begin{eqnarray}
Q_0&=& \oint c\Big(T_X + T_{\phi} + T_{\rm L} - s\, \beta\del\gamma -
(s-1) \del\beta\, \gamma -b\del c\Big)\ ,\nonumber\\
Q_1&=& \oint \gamma\Big( (\del\phi)^s + \ft12 s^2 (\del\phi)^{s-2}\beta\del
\gamma - 2(\del\phi)^{s-2} T_{\rm L} - (s-2)^2 (\del\phi)^{s-4}\,
T_{\rm L}\, \beta\del\gamma\Big)\ .\label{w2s2}
\end{eqnarray}

     Having obtained the classical BRST operators for $W^{\rm M}_{2,s}
\otimes W^{\rm L}_2$ and $W^{\rm M}_{2,s}\otimes W^{\rm L}_{2,s'}$, we now
turn our attention to their quantisation.   For generic values of $s$ and
$s'$, the associated algebras do not close except at discrete sets of values
of the central charges.  Thus it is not possible to construct abstract
quantum BRST operators for generic $s$ and $s'$.  However, when $s$ and $s'$
take their values in the set $\{2, 3, 4, 6\}$ the corresponding algebras
close for generic values of central charge.  Thus at the quantum level, in
addition to the previously constructed critical $W_{2,s}$ BRST operators and
non-critical BRST operators with $s=s'$, there are four more non-critical
BRST operators that can be possibly constructed abstractly, namely,
$W^{\rm M}_{2,4}\otimes W^{\rm L}_2$, $W^{\rm M}_{2,6}\otimes W^{\rm L}_2$,
$W^{\rm M}_{2,6}\otimes W^{\rm L}_{2,3}$ and
$W^{\rm M}_{2,6}\otimes W^{\rm L}_{2,4}$.
In this paper, we construct abstract quantum BRST operators for $W_{2,s}^{\rm
M}\otimes W^{\rm L}_2$ with $s=4, 6$.  These two BRST operators turn out to
be surprisingly simple, with a graded structure $Q=Q_0+Q_1$ where
$Q_0^2=Q_1^2 =\{Q_0, Q_1\} = 0$.

      For the case $s=4$, we write down the most general possible structure,
subject to the requirement that $\{Q, b\}$ gives rise to the standard total
matter-plus-ghost energy-momentum tensor, and solve for the coefficients by
demanding nilpotence.  We find that the BRST operator can be nilpotent
provided that the central charges for the matter and Liouville sectors are
$c_{\rm M}=196$ and $c_{\rm L}=-24$.  The BRST operator has two free
parameters, which correspond to the freedom to perform canonical
transformations on the ghost fields.  Choosing the parameters suitably, we
find that the BRST operator can be written in a graded form, given by
\begin{eqnarray}
Q_0&=& \oint c\Big ( T_{\rm M} + T_{\rm L} - 4\beta\del\gamma -3
\del\beta\, \gamma - b\del c\Big)\ ,\nonumber\\
Q_1&=& \oint \gamma\Big( \sqrt{\ft{130594}{11}}\, W_{\rm M} +
\ft{113}{132} T_{\rm M}^2
+\ft{445}{88} \del^2 T_{\rm M} -\ft{167}{22} T_{\rm M}\, \beta\del\gamma
-\ft{1169}{44}\beta\del^3\gamma + \ft{1169}{66} \del^2\beta\, \del\gamma
\Big)\ .\label{w242ab}
\end{eqnarray}

    It is interesting to compare this BRST operator with the usual
non-critical BRST operator for $W^{\rm M}_{2,4}\otimes W^{\rm L}_{2,4}$
at $c_{\rm L}=-24$.  At this particular value of central charge, the
$W_{2,4}$ algebra becomes singular.   However, we find that one can rescale
the spin--4 current such that no coefficients diverge in the OPE
$W(z)\, W(w)$, which is now given by
\begin{equation}
W(z)\, W(w)\sim {2 TW -\ft13 \del^2 W \over (z-w)^2} + {\del(TW) -
\ft16 \del^3 W \over z-w }\ .\label{w24null}
\end{equation}
It should be emphasised that the algebra generated by $T$ and $W$ is
consistent and satisfies the Jacobi identity.  We find that the usual
non-critical BRST operator for $W^{\rm M}_{2,4}\otimes W^{\rm L}_{2,4}$ at
$c_{\rm L}=-24$ is given by
\begin{equation}
Q=Q_0 + Q_1 -\oint \gamma\Big(\ft{167}{22}\, W_{\rm L} +
\ft{27889}{484}\, W_{\rm L}\,b\del\gamma\Big)\ ,\label{w244brst}
\end{equation}
where the first two terms are precisely given by Eqn.\ (\ref{w242ab}).  Now
it is easy to understand why the BRST operator (\ref{w242ab}) for $W^{\rm
M}_{2,4}\otimes W^{\rm L}_2$ exists.  It follows from the OPE
(\ref{w24null}) that we can consistently set $W_{\rm L}$ to zero since it
appears in every term on the right hand side of the OPE. Thus the BRST
operator (\ref{w244brst}) will continue to be nilpotent if $W_{\rm L}$ is
set to zero.

   There are three other values of the central charge for which the
$W_{2,4}$ algebra becomes singular, namely $c=\ft12, -\ft{68}{7}$ and
$-\ft{22}{5}$.  For these values, we can also rescale the spin--4 current
such that its OPE with itself has no divergent coefficients.  In each case,
however, we find that the result fails to satisfy the Jacobi identity.  Thus
we expect that there is no non-critical BRST operator for $W^{\rm
M}_{2,4}\otimes W^{\rm L}_{2,4}$ when $c_{\rm L}$ is equal to any of these
three values.   By contrast, as we have shown, when $c=-24$ the $W_{2,4}$
algebra is consistent, although degenerate, and we have constructed the
associated abstract non-critical BRST operator.  The existence of the
algebra at $c=-24$ was not seen in the previous results in Refs.\
\cite{kw,bfknrv}.  This can be understood as follows.  For generic values of
central charge, the $W_{2,4}$ algebra can be expressed in the simple form
\cite{kw,bfknrv}
\begin{equation}
W\, \circ\, W = \ft{c}{4} {\hbox{ \bf 1}} + q\, W\ ,\label{w24sim}
\end{equation}
where
\begin{equation}
q^2 = {54 (c+24) (c^2-172 c + 196) \over (2c-1) (7c+68) (5c+22)}\ .
\end{equation}
In the abbreviated notation of Eqn.\ (\ref{w24sim}), the right-hand side
denotes the identity operator and the spin--4 current and all their
descendants.   This expression is however misleading for special values of
central charge, since some of the descendant terms have additional central
charge dependent factors that can diverge at the special values.  In
particular,  the $W$ descendant $TW-\ft16 \del^2 W$ has a
coefficient factor $1/(c+24)$.  We can then rescale the $W$ current such
that $W\longrightarrow W/\sqrt{c+24}$, and thereby obtain the $W_{2,4}$
algebra at $c=-24$, as given by (\ref{w24null}).

    The degeneration of the $W_{2,4}$ algebra at $c=-24$ is analogous to the
degeneration of the $W_3$ algebra at $c=-\ft{22}{5}$.  In this latter case,
one can again rescale the spin--3 current, thereby obtaining the OPE
\begin{equation}
W(z)\, W(w)\sim {2 (T^2-\ft3{10}\del^2 T)\over (z-w)^2} +
{\del T^2 -\ft3{10} \del^3 T\over z-w}\ .\label{w3null}
\end{equation}
It is easy to verify that $T$ and $W$ define a consistent algebra which
satisfies the Jacobi identity.   We have verified that it can be used to
construct the non-critical BRST operator for $W^{\rm M}_3\otimes
W^{\rm L}_3$ with $c_{\rm L} = -\ft{22}5$.  However, in this case one cannot
obtain a non-critical BRST operator for $W^{\rm M}_3\otimes W^{\rm L}_2$ by
setting the current $W_{\rm L}$ to zero, since the right hand side of
Eqn.\ (\ref{w3null}) does not vanish.   In fact we have verified that no
$W^{\rm M}_3\otimes W^{\rm L}_2$ BRST operator exists.

    For the case of $s=6$, we make the assumption that the BRST operator has
the same graded structure, and we find that it is nilpotent when $c_{\rm
M}=390$, $c_{\rm L}=-2$ or $c_{\rm M}=\ft{1450}3$, $c_{\rm L}=-\ft{286}3$.
For the first case, the BRST operator is given by
\begin{eqnarray}
Q_0&=&\oint c\Big(T_{\rm M} + T_{\rm L }-
6\beta\del \gamma - 5\del\beta\, \gamma -
b\del c\Big)\ ,\nonumber\\
Q_1&=&\oint \gamma \Big( \st{2448} \sqrt{\ft{41149461318}{13}}\, W_{\rm M} +
\st{4282}\, T_{\rm M}^3 +
\ft{1390837}{13}\, \del^2 T_{\rm M}\, T_{\rm M} +
\ft{1038100}{13}\, \del T_{\rm M}\, \del T_{\rm M} \nonumber\\
&&+\ft{6815257}{39}\, \del ^4 T_{\rm M}  -
\ft{1032462}{13}\, T_{\rm M}^2\,\beta\del\gamma +
\ft{4301925}{13}\, \del^2 T_{\rm M}\, \beta\del\gamma\label{w26brst}\\
&&+\ft{16634110}{13}\, \del T_{\rm M}\, \del \beta\, \del\gamma
+\ft{6653644}{13}\, T_{\rm M}\,\del^2\beta\, \del\gamma
-\ft{9980466}{13}\, T_{\rm M}\,\beta\del^3\gamma
-\st{1433975}\,\del^4\beta\, \del\gamma\nonumber\\
&& +  \st{2581155}\, \del^2\beta\del^3\gamma
-\st{1433975}\,\beta\del^5\gamma-
\st{1720770}\, \del\beta\, \beta\del^2\gamma\, \del\gamma\Big)\ .\nonumber
\end{eqnarray}
For the latter case, $Q_0$ takes an identical form to $Q_0$ in Eqn.\
(\ref{w26brst}), and the $Q_1$ operator is given by
\begin{eqnarray}
Q_1&=&\oint\gamma \Big ( \st{136}\sqrt{\ft{4609647394209}{483799}} W_{\rm M}
-\ft{3557034}{483799} T^3_{\rm M} -\ft{89012250}{483799} (\del T_{\rm M})^2
-\ft{98014833}{483799} \del^2 T_{\rm M}\, T_{\rm M}\nonumber\\
&& -\ft{103809276}{483799} \del^4 T_{\rm M}
+\st{1746} T_{\rm M}\, \del \beta\del^3\gamma  + \st{162} (T_{\rm
M})^2\,\beta\del\gamma -\st{1224} T_{\rm M}\,\del^2\beta\, \del\gamma
\label{w26brst1}\\
&& -
\st{1215} \del^2T_{\rm M}\, \beta\del\gamma +\st{2645} \beta\del^5\gamma
+ \st{6675 }\del^2\beta\, \del^3\gamma +
\st{4240} \del^4\beta\, \del\gamma -\st{1350}
\del\beta\,\beta\del^2\gamma\,\del\gamma\Big)\ .\nonumber
\end{eqnarray}

    The existence of the BRST operators (\ref{w26brst}) and (\ref{w26brst1})
can be understood from the fact that the $W_{2,6}$ algebra becomes
degenerate when $c=-2$ and $c=-\ft{286}3$, analogous to the case of
$W_{2,4}$ at $c=-24$.   At $c=-2$, we find that the OPE of the spin--6
current with itself gives rise only to a null current:
\begin{equation}
W(z)\, W(w) \sim {2\Lambda\over (z-w)^2} + {\del \Lambda
\over z-w}\ ,\label{w26null}
\end{equation}
where $\Lambda= T^2 W -\ft59 T\del^2 W +
\ft9{19} \del T\, \del W -\ft{17}6 \del^2 T\, W +\ft{1}{36} \del^4 W$.
At $c=-\ft{286}3$, we find that the OPE of the spin--6 current $W$ with itself
contains only $W$ and its descendants:
\begin{eqnarray}
&&W(z)\, W(w) \sim\nonumber\\
&& {W\over (z-w)^6 }+\ft12{\del W\over (z-w)^5}
+\ft7{52} {\del^2 W\over (z-w)^4} +\ft1{39} {\del^3 W\over (z-w)^3} +
\ft{1}{260} {\del^4 W\over (z-w)^2} +
\ft{1}{2080} {\del^5 W\over z-w} \label{w26null1}\\
&& -\ft9{35}{\Lambda_1 \over (z-w)^4} -
\ft{9}{70}{\del \Lambda_1 \over (z-w)^3}
-\ft{81}{2380} {\del^2\Lambda_1 \over (z-w)^2}
-\ft{3}{476}{\del^3\Lambda_1 \over z-w} +
\ft{27}{700} {\Lambda_2 \over (z-w)^2} +
\ft{27}{1400} {\del\Lambda_2 \over z-w}\ ,\nonumber
\end{eqnarray}
where $\Lambda_1 = TW -\ft3{26} \del^2 W$ and $\Lambda_2 =
T^2W -\ft{35}{153}T\del^2 W -\ft{2}{153} \del T\, \del W -\ft{29}{102}
\del^2T\, W + \ft{7}{612}\del^4W$.   Thus in the BRST operator
for $W^{\rm M}_{2,6}\otimes W^{\rm L}_{2,s}$, we can consistently set the
spin--6 Liouville current $W_{\rm L}$ to zero when $c_{\rm L}=-2$ and when
$c_{\rm L}=-\ft{286}3$, giving rise to the BRST operators (\ref{w26brst})
and (\ref{w26brst1}) respectively.

    Having obtained the non-critical abstract BRST operators for $W^{\rm
M}_{2,s}\otimes W^{\rm L}_2$ with $s=4$ and 6, it is interesting to
investigate the implications for the corresponding string theories.  We can
realise the matter $W^{\rm M}_{2,s}$ currents in terms of free scalars, with
the classical terms given by Eqn.\ (\ref{w2sclreal}).  For $s=4$ and both
the above cases for $s=6$, it turns out that the central charge required by
criticality implies that the central charge of the effective energy-momentum
tensor $T_X$ is $c_X=26$. Note that the Liouville current appears only in
$Q_0$ in the abstract BRST operators (\ref{w242ab}) and
(\ref{w26brst}), whilst in the BRST operator in terms of scalar
realisations, the Liouville current appears both in the $Q_0$ and $Q_1$
operators.   We have verified for $s=4$ that in terms of the free-scalar
realisation, the BRST operator (\ref{w2s2}) with its quantum corrections can
be converted into the BRST operator (\ref{w242ab}) by a local canonical
field redefinition.

     Now let us turn our attention to the quantisation of the non-critical
BRST operators for $W^{\rm M}_{2,s}\otimes W^{\rm L}_2$ with $s$ taking
values other than 4 and 6.  In these cases, we have to start with
a specific realisation for the classical $W_{2,s}$ algebra, since at the
quantum level it fails to satisfy the Jacobi identify for generic values of
central charge except when $s=4, 6$.  We find that for these non-critical
BRST operators, just as in the case of the critical $W_{2,s}$ BRST operators
discussed previously \cite{w2s,lptwx}, nilpotence can be achieved at the
quantum level even when the corresponding quantum $W_{2,s}$ algebra does not
exist. The classical BRST operators for $W^{\rm M}_{2,s}\otimes W^{\rm L}_2$
in terms of scalar realisations are given by Eqn.\ (\ref{w2s2}).  To
quantise the BRST operators, we can add all possible $\hbar$--dependent
quantum corrections and demand nilpotence.  We have studied the case of
$s=5$ in detail.  Assuming the graded form of the BRST operator is preserved
at the quantum level, we find by explicit computation that there is one
solution. It is given by
\begin{eqnarray}
Q_0&=&\oint c\Big( T_X -\ft12 (\del\phi)^2 +T_{\rm L}-6\beta\del\gamma -
5 \del\beta\,
\gamma -b\del c\Big)\ ,\nonumber\\
Q_1&=&\oint \gamma\Big( (\del\phi)^5 - 2(\del\phi)^3\,T_{\rm L}
+\ft{25}{2} (\del\phi)^3\beta\, \del\gamma - 9 \del\phi\, T_{\rm L}
\beta\del\gamma -\ft{65}{2\sqrt{2}} \del^2\phi\, (\del\phi)^3\nonumber\\
&&+\ft{165}{4} (\del^2\phi)^2
\del\phi +\ft{175}{8}\del^3\phi\, (\del\phi)^2 -
\ft{265}{8\sqrt{2}} \del^3\phi\, \del^2\phi
-\ft{55}{4\sqrt{2}}\del^4\phi\, \del\phi +\ft{53}{32}\del^5\phi\nonumber\\
&& +\ft{9}{\sqrt{2}} (\del\phi)^2\, \del T_{\rm L} +\ft{21}{\sqrt{2}}
\del^2\phi\, \del\phi\, T_{\rm L} -\ft{9}{4} \del\phi\, \del^2 T_{\rm L}
-\ft{27}{4} \del^2\phi\, \del T_{\rm L}
-\ft{19}{4} \del^3\phi\, T_{\rm L}\label{w252}\\
&& +\ft{\sqrt{2}}{16} \del^3 T_{\rm L} -\ft{45}{2\sqrt{2}} (\del\phi)^2
\del\beta\, \del\gamma  +\ft{75}{2\sqrt{2}} (\del\phi)^2\beta\del^2\gamma
-\ft{45}{2} \del\phi\, \del\beta\, \del^2\gamma
-\ft{25}{4}\del^2\phi\, \beta\del^2\gamma\nonumber\\
&& -\ft{65}{8} \del^2\phi\, \del \beta\, \del\gamma
-\ft{21}{2\sqrt{2}} T_{\rm L}\beta\del^2\gamma -\ft{15}{2\sqrt{2}}
\del T_{\rm L}\, \beta\del\gamma -\ft{65}{8\sqrt{2}} \beta\del^4\gamma
 -\ft{65}{4\sqrt{2}}\del\beta\,\del^3\gamma\Big)\ .\nonumber
\end{eqnarray}
The central charge for the effective energy-momentum tensor is again
$c_X=26$.   We expect that the classical BRST operators (\ref{w2s2}) for
$W^{\rm M}_{2,s}\otimes W^{\rm L}_2$ are quantisable for all values of $s$,
leading in each case to a string with an effective 26 dimensional spacetime.

     Next, we consider the non-critical BRST operator for the $W^{\rm
M}_{2,4}\otimes W^{\rm L}_4$ system.  Owing to the complexity of the $W_4$
algebra, we have only considered this BRST operator at the classical level.
Classically, the $W_4$ currents in the Liouville sector, namely $T_{\rm L}$,
$W_{\rm L}$ and $U_{\rm L}$ with spins 2, 3 and 4 respectively, have the
following OPEs amongst the higher-spin currents:
\begin{eqnarray}
W_{\rm L}(z) W_{\rm L}(w) &\sim& {\ft13 T_{\rm L}^2 +2 U_{\rm L} \over
(z-w)^2} + {\ft16 \del(T_{\rm L})^2 + \del U_{\rm L} \over z-w}\ , \nonumber\\
W_{\rm L}(z) U_{\rm L}(w) &\sim& {\ft53 T_{\rm L} W_{\rm L} \over
(z-w)^2} + {\ft23 \del(T_{\rm L} W_{\rm L})\over z-w}\ , \label{clw4}\\
U_{\rm L}(z) U_{\rm L}(w) &\sim& {\ft29 T_{\rm L}^3 +2 (W_{\rm L})^2 \over
(z-w)^2} + {\ft19 \del(T_{\rm L})^3 + \del (W_{\rm L})^2 \over z-w}\ .\nonumber
\end{eqnarray}
In terms of a free-scalar realisation of the $W_{2,4}$ algebra of the form
(\ref{w2sclreal}), we find that the classical $W^{\rm M}_{2,4}\otimes W^{\rm
L}_4$ non-critical BRST operator takes the graded form
\begin{eqnarray}
Q_0 &=& \oint c\Big(T_X + T_\phi + T_{\rm L} -4\beta\del\gamma -3
\del\beta\, \gamma -b\del c\Big)\ ,\nonumber\\
Q_1 &=& \oint\gamma \Big( (\del\phi)^4 + 8 (\del\phi)^2 \beta\del\gamma -
\ft43 (\del\phi)^2 \, T_{\rm L}-\ft43 U_{\rm L} \label{w244clbrst}\\
&&+\ft29 (T_{\rm L})^2 -\ft{4\sqrt{2}}{3}\del\phi\, W_{\rm L} -
\ft{16}{9} T_{\rm L} \beta\del\gamma \Big)\ .\nonumber
\end{eqnarray}

     To summarise, we have studied in this paper the construction of the
non-critical BRST operators whose matter and Liouville currents generate
two different $W$ algebras.   At the classical level, we obtained such BRST
operators for $W^{\rm M}_{2,s}\otimes W^{\rm L}_{2,s'}$.  The construction
is possible for $s=s'$ or $s\ge s'+2$.  We also obtained the classical BRST
operator for $W^{\rm M}_{2,4} \otimes W^{\rm L}_4$.  We used free-scalar
realisations for the matter currents, and the corresponding BRST operators
have a graded form.  It would be interesting to extend these results to the
quantum level.  However, in general the computations become unmanageably
complicated.  We have explicitly constructed the BRST operators for $W^{\rm
M}_{2,s} \otimes W^{\rm L}_2$ with $s=4,5,6$ at the quantum level.  For the
first and the third cases, we obtained the BRST operators in terms of
abstract matter and Liouville currents.  This is not possible when $s=5$
since the $W_{2,5}$ algebra does not exist at the quantum level for the
value of central charge required by the criticality.   We started with the
classical BRST operator for $W^{\rm M}_{2,5}\otimes W^{\rm L}_2$ with the
matter currents realised in terms of free scalars, and obtained the
nilpotent quantum BRST operator by adding $\hbar$--dependent
corrections.   This construction should presumably work for all values of
$s$.  In terms of the free-scalar realisations for the matter currents, the
central charge for the effective energy-momentum tensor $T_X$ is $c_X=26$
for all the cases $s=4,5,6$.  We expect that it will be true also for higher
values of $s$.

      It is of interest to extend our results to more general classes of $W$
algebras.   The classification of possible BRST operators that can be built
becomes equally as interesting as the classification of the $W$ algebras
themselves.   For any algebra $W$, we expect that there should exist a
critical BRST operator and a non-critical BRST operator for $W^{\rm M}
\otimes W^{\rm L}$ at the classical level.  If the algebra exists also at
the quantum level for generic values of central charge, we expect that we
can obtain the quantum BRST operator abstractly; otherwise, the quantisation
has to be carried out in terms of a specific realisation.   If one allows
two different $W$ algebras for the matter and Liouville sectors, the
possibilities become more numerous.   Although we have constructed the
classical BRST operators for a large class of $W$ algebras, the
quantisability of these BRST operator remains to be understood.  We obtained
the full quantum BRST operators for $W^{\rm M}_{2,s}\otimes W^{\rm L}_2$
with $s=4,5,6$.  For the cases $s=4$ and 6, the existence of the quantum
BRST operators can be understood as special cases of $W^{\rm M}_{2,s}\otimes
W^{\rm L}_{2,s}$ where the spin--4 or spin--6 Liouville currents can be
consistently set to zero at the special values of central charge that arose
in our construction.   For the $s=5$ case, on the other hand, there is no
such explanation.   In view of the complexity of the computations, the
understanding of the general structure of the BRST operators requires a less
empirical approach.

\section*{Acknowledgements}

     We are grateful to R.\ Blumenhagen and H. Kausch for discussions.

\end{document}